\documentclass[aps,prb,10pt,twocolumn,superscriptaddress,longbibliography]{revtex4-2}
\usepackage{graphicx}
\usepackage{dcolumn}
\usepackage{bm}
\usepackage{color}
\usepackage{amsmath,amssymb}
\usepackage{physics}
\usepackage{booktabs}
\usepackage{hyperref}
\hypersetup{colorlinks=true,breaklinks,linkcolor=blue,urlcolor=blue,citecolor=blue}
\begin{document}
\newcommand{\Ns}{N_{\text{s}}}
\newcommand{\bvec}[1]{\mbox{\boldmath$#1$}}
\def\vec#1{\boldsymbol #1}

\title{
Revisiting spin Hamiltonian parameters in a Kitaev material
via Bayesian optimization of magnetization curves
}
\author{Takahiro Misawa}
\affiliation{Institute for Solid State Physics, University of Tokyo, Kashiwa-no-ha, Kashiwa-shi, Chiba, 277-8581, Japan}
\author{Ryo Tamura}
\affiliation{Center for Basic Research on Materials, National Institute for Materials Science, Namiki, Tsukuba-shi, Ibaraki, 305-0044, Japan}
\affiliation{Graduate School of Frontier Sciences, The University of Tokyo, Kashiwa-no-ha, Kashiwa-shi, Chiba, 277-8568, Japan}
\author{Kazuyoshi Yoshimi}
\affiliation{Institute for Solid State Physics, University of Tokyo, Kashiwa-no-ha, Kashiwa-shi, Chiba, 277-8581, Japan}
\author{Youhei Yamaji}
\affiliation{Research Center for Materials Nanoarchitectonics (MANA),\\ National Institute for Materials Science, Namiki, Tsukuba-shi, Ibaraki, 305-0044, Japan}

\begin{abstract}
Determining the spin Hamiltonian of a magnetic compound is crucial for understanding its magnetic properties.
A standard approach is to derive model parameters from $ab$ $initio$ calculations based on the
crystal structure. However, the resulting Hamiltonian can depend sensitively on methodological details of the $ab$
$initio$ procedure. This issue is particularly evident in $\alpha$-RuCl$_3$, a candidate Kitaev material. Here, we
present an alternative, data-driven approach to determine the spin Hamiltonian parameters of $\alpha$-RuCl$_3$ by Bayesian
optimization of experimental magnetization curves along the $b$- and $c$-axis directions. We optimize five parameters,
namely the Kitaev interaction $K$, off-diagonal interactions $\Gamma$ and $\Gamma'$, the Heisenberg interaction $J$, and
the $c$-axis $g$-factor $g_c$. The parameter set that minimizes the cost function is
$(K,\Gamma,\Gamma',J,g_c)=(-6.0,\,7.5,\,-0.3,\,-1.75,\,2.3)$, where
the exchange couplings are in meV. We find that the cost function is insensitive to
the absolute value of the Kitaev coupling $K$. Thus, the magnetization data alone do not
determine its energy scale. The cost function also depends only weakly on $\Gamma'$ and $J$,
while the optimization favors a large positive $\Gamma$. By computing the static spin structure factor,
magnetic susceptibility, and specific heat, we show that these quantities favor the large-$\Gamma$ scenario over
the small-$g_c$ scenario and that the parameter set that minimizes the cost function yields good
agreement with the experimental structure factor and susceptibility, and qualitative agreement with the specific heat. The combination of Bayesian optimization and accurate low-energy solvers provides an effective
approach for determining parameters of spin Hamiltonians. This methodology opens a systematic route to determining
spin Hamiltonians in quantum magnets from experimental data.
\end{abstract}
\maketitle

\section{Introduction}
Constructing microscopic models that describe the physical properties of materials has played an essential role
in understanding various emergent phenomena in condensed matter physics. Famous examples include the Heisenberg model
for magnetism and the Hubbard model for strongly correlated electron systems~\cite{Auerbach1994,Fazekas1999,
Tasaki2020}. Detailed analysis of these
celebrated effective models and their extensions has provided deep insights into emergent phenomena such as
magnetism, metal-insulator transitions, and superconductivity~\cite{Imada_RMP1998,Lee_RMP2006}.

In recent years, considerable effort has been devoted to deriving quantitative low-energy effective Hamiltonians based
on $ab$ $initio$ calculations. One prominent approach is the $ab$ $initio$ downfolding method~\cite{Imada_JPSJ2010}, which begins
with calculating the electronic band structure based on lattice structures of solids. Using the constrained
random phase approximation (cRPA)~\cite{Aryasetiawan_PRB2004}, one can then evaluate the screened Coulomb interactions
within the low-energy degrees of freedom.
This procedure enables the construction of material-specific Hubbard-type Hamiltonians, whose parameters are
fully determined by the $ab$ $initio$ procedure. The method has been successfully applied to a
wide variety of strongly correlated systems~\cite{Nakamura_PRB2009,Nohara_JPSJ2011,
Wehling_PRL2011,Saioglu_PRL2012,
Nomura_PRB2012,Vaugier_PRB2012,
Arita_PRL2012,Nilson_PRB2013,
Hansmann_JPS2013,Yamaji_PRL2014,
Okamoto_PRB2014,Amadon_PRB2014,
Kim_PRB2016_2,Seth_PRL2017,
Moree_PRB2018,Nomura_PRB2019,
Hirayama_PRB2020}, including iron-based superconductors~\cite{Nakamura_JPSJ2008,Miyake_JPSJ2010b,
Misawa_JPSJ2011,Misawa2014_FeAs_NCom,
Hirayama2015_FeTe_JPSJ}, cuprates~\cite{Hirayama_PRB2018,Tadano_PRB2019,
Hirayama_PRB2019}, and molecular solids~\cite{Nakamura_JPSJ2009,Nakamura_PRB2012,
Shinaoka_JPSJ2012,Misawa2020_dmit_PRR,
Yoshimi2021_dmitfull_PRR,Ido2022_dmit_npjQM,
Ohki2023_alpha_PRB,Yoshimi2023_TM_PRL,
Kawamura2024_EDOTTFI_PRL,Itoi2024_PressureTM_PRR,
Kato2025_DMET_PRB}.

Efforts have also been made to obtain spin Hamiltonians for magnetic materials. One standard approach
is to first derive a Hubbard-type Hamiltonian via the $ab$ $initio$ downfolding and then perform
a strong-coupling expansion to obtain the effective spin models~\cite{Nakamura_PRB2009,Yamaji_PRL2014,
Solovyev_PRB2015,Eichstaedt_PRB2019,
Villanova_PRR2023}. While this approach has succeeded in
analyzing magnetic properties in several materials, it is difficult to determine a systematic criterion for
deciding the necessary order of perturbative expansion. Another well-established method is based on the Lichtenstein
formula~\cite{Liechtenstein_JMMM1987}, which evaluates exchange interactions between classical spins
by calculating the energy variation upon infinitesimal
rotation of magnetic moments around a given magnetic configuration. Although this method has been widely
used, the validity of the resulting classical spin models remains unclear for materials with strong
quantum fluctuations. Extensions to Dzyaloshinskii-Moriya interactions have been proposed~\cite{Udvardi_PRB2003,Katsnelson_PRB2010,
Szilva_RMP2023}, but the framework is not readily
applicable to general non-Heisenberg terms such as the Kitaev and off-diagonal $\Gamma$ interactions, which are
essential in spin-orbit-coupled materials.

Among several quantum magnets, $\alpha$-RuCl$_3$~\cite{Plumb_PRB2014,Sears_PRB2015} is a prototypical example where the determination of the spin
Hamiltonian remains controversial. This compound has attracted considerable attention as a promising candidate for realizing
the Kitaev quantum spin liquid~\cite{Kitaev_AnnPhys2006,Jackeli_PRL2009,
Motome_JPSJ2020,3m4m-3v59}. In particular, the reported half-integer quantization of the thermal Hall
conductivity~\cite{Kasahara_Nature2018,Yokoi_Science2021} has been interpreted as evidence for Majorana fermions,
although this interpretation remains under active
debate~\cite{Czajka_NPhy2021,bruin2022robustness}.
Estimated spin-Hamiltonian parameters for $\alpha$-RuCl$_3$
(including the Kitaev interaction $K$, off-diagonal interactions $\Gamma$ and
$\Gamma'$, and Heisenberg interaction $J$) show substantial method dependence~\cite{Kim_PRB2016,Winter_PRB2016,
Yadav_SR2016,Ran_PRL2017,
Wang2017,Winter2018,
Suzuki_PRB2018,Ozel2019,
Eichstaedt_PRB2019,Laurell_npjQM2020,
Maksimov_PRR2020,Sears2020,
Li_NC2021,Suzuki2021RIXS,
PhysRevResearch.4.L022061,Moller_PRB2025}. These estimates include both theoretical and
experimental studies. Complementary experimental constraints on the sign of the Kitaev interaction and the role
of anisotropic exchanges have also been discussed extensively~\cite{Winter2018,Sears2020,
Suzuki2021RIXS}, highlighting the ambiguity inherent in conventional approaches.
Detailed comparisons of the proposed parameter sets have been performed in Refs.~\cite{Maksimov_PRR2020,Moller_PRB2025}.

To re-examine these magnetic exchange couplings in Kitaev materials, previous studies have analyzed the temperature
dependence of the specific heat~\cite{Li_NC2021} and spin excitation spectra~\cite{Suzuki2021RIXS}, which appear to directly reflect the
relevant energy scales. However, both observables present challenges: the former exhibits strong size sensitivity near
phase transitions, complicating model construction, whereas continua in the latter can be difficult to identify
in samples with defects or stacking disorder. The magnetic-field dependence of the magnetization provides a
complementary route to estimate the exchange couplings. While a divergence may appear in its derivative
in the thermodynamic limit, the magnetization process of Kitaev materials is continuous with respect to
the magnetic field~\cite{PhysRevB.82.064412,Johnson_PRB2015,
PhysRevB.97.075126}. This smoothness mitigates the finite-size sensitivity that complicates analyses based on thermodynamic
quantities. The magnetization process has been used as a
fingerprint of various frustrated magnets~\cite{miyahara2003theory,takigawa2010magnetization,
yoshida2022frustrated}.
Among Kitaev materials, the magnetization of $\alpha$-RuCl$_3$ is particularly informative, exhibiting a field-induced intermediate phase and
pronounced field-direction anisotropy~\cite{Johnson_PRB2015,PhysRevB.97.075126}. In contrast, the related material Na$_2$IrO$_3$ exhibits comparatively featureless magnetization curves~\cite{PhysRevB.82.064412}.

In this paper, we present an alternative approach for estimating the spin Hamiltonian of $\alpha$-RuCl$_3$
by solving an inverse problem using experimental observables. Specifically, we combine exact diagonalization using $\mathcal{H}\Phi$~\cite{Kawamura_CPC2017,Ido_CPC2024,
HPhi_URL,HPhi_github}
with Bayesian optimization using PHYSBO~\cite{Motoyama_CPC2022,PHYSBO_URL,
PHYSBO_github} to determine the optimal parameter set that reproduces the experimental
magnetization process. We focus on five key parameters, namely $K$, $\Gamma$, $\Gamma'$, $J$, and the
$c$-axis $g$-factor $g_c$. The optimal parameter set is $K=-6.0$, $\Gamma=7.5$, $\Gamma^{\prime}=-0.3$, $J=-1.75$ (in meV), and
$g_{c}=2.3$. We demonstrate that the best parameter set not only reproduces the experimental magnetization curves
but also yields good agreement with the static spin structure factor and magnetic susceptibility, and
qualitative consistency with the specific heat. By analyzing the cost-function landscape using Gaussian process regression,
we find that the magnetization process is remarkably insensitive to the Kitaev coupling $K$, so
that the absolute energy scale cannot be determined from magnetization data alone. Our approach offers
a general framework for determining effective spin models based on experimental data, complementing the limitations
of traditional $ab$ $initio$ methods.

The remainder of this paper is organized as follows. Section~\ref{sec:model} introduces the spin Hamiltonian for
$\alpha$-RuCl$_{3}$ and reviews previous parameter estimates. Section~\ref{sec:method} describes
the Bayesian-optimization procedure and exact-diagonalization setup. Section~\ref{sec:results}
presents the optimization results, analyzes the cost-function landscape, and examines the spin structure factor, magnetic
susceptibility, and specific heat. Section~\ref{sec:summary} summarizes the main conclusions.

\section{Effective Spin Hamiltonian}
\label{sec:model}

\begin{figure}[t]
\begin{center}
\includegraphics[width=0.4\textwidth]{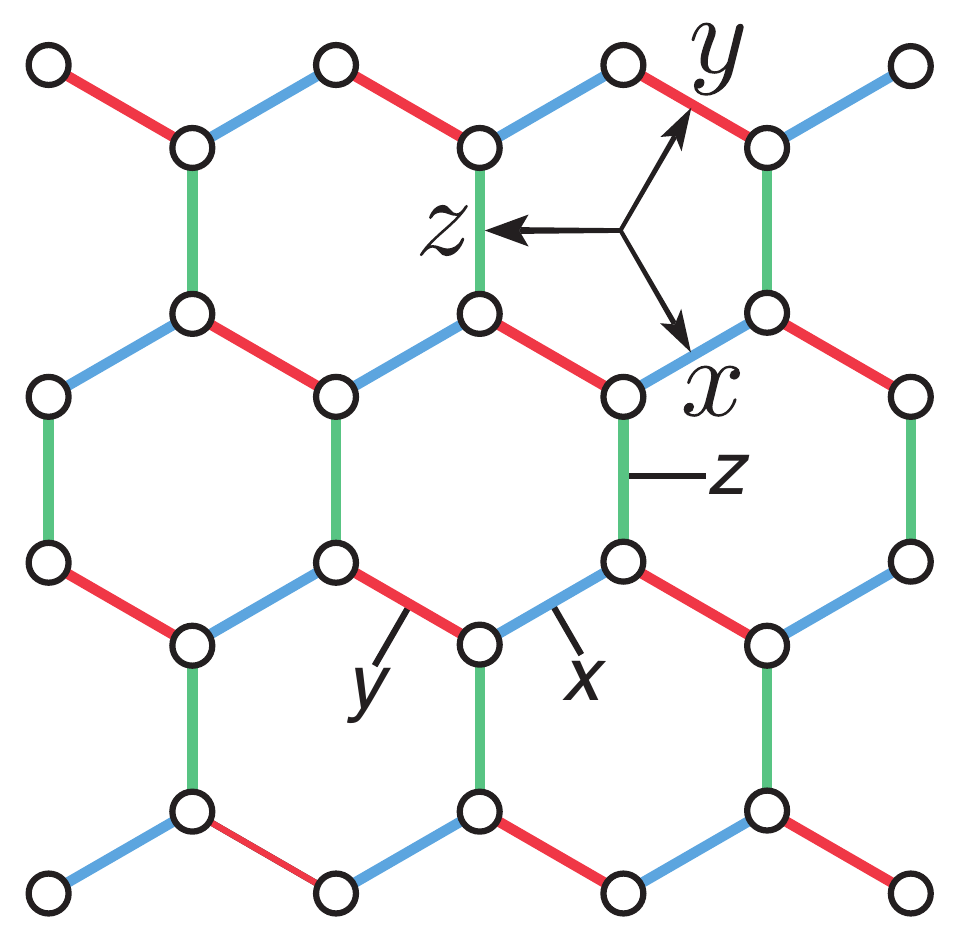}
\end{center}
\caption{ Honeycomb network of Ru atoms in $\alpha$-RuCl$_3$ with the $R\overline{3}$ structure. Edge-sharing chlorine octahedra
surround the Ru sites, shown by open circles.
The blue, red, and green solid lines represent the three inequivalent nearest-neighbor bonds
(the \textsf{\textit{x}}, \textsf{\textit{y}}, and \textsf{\textit{z}} bonds, respectively) on which the bond-dependent exchange couplings act.
The black arrows labeled $x$, $y$, and $z$ indicate the cubic spin axes defining
the spin components in Eq.~(\ref{eq:ham}). The bond labels are written in the sans-serif
italic font to distinguish them from the spin axes.
}
\label{fig:honeycomb}
\end{figure}

In the present paper, we focus on the low-energy degrees of freedom of $\alpha$-RuCl$_3$ that
are represented by spin-1/2 objects on the Ru sites (Fig.~\ref{fig:honeycomb}). For simplicity, we assume a
trigonal symmetry, $R\overline{3}$, of the crystal structure at low temperatures~\cite{Cao_PRB2016,Kim_PRB2024},
instead of monoclinic $C2/m$~\cite{Johnson_PRB2015}.
Among low-energy Hamiltonians consistent with the crystal symmetry~\cite{Rau_PRL2014},
we consider the following spin Hamiltonian under a
magnetic field:
\begin{align}
\mathcal{H} =\sum_{\langle i,j \rangle_\gamma}\sum_{\alpha,\beta = x,y,z}
J_{\alpha\beta}^{\gamma} S_i^{\alpha} S_j^{\beta}
-g_{\eta}\mu_{\rm B}\sum_{i}\vec{h}_{\eta}\cdot\vec{S}_i,
\label{eq:ham}
\end{align}
where $\langle i,j\rangle_\gamma$ denotes the nearest-neighbor pair on the $\gamma$-bond. The index $\eta =
b, c$ labels the magnetic-field direction, with $\vec{h}_b \propto [1,-1,0]$ for $H\parallel b$ and $\vec{h}_c
\propto [1,1,1]$ for $H\parallel c$. The corresponding $g$-factor $g_\eta$ depends on the field direction. In
this paper, we fix $g_a = g_b = g_{ab} = 2.5$~\cite{Kubota_PRB2015,Sears_PRB2015,
Ponomaryov_PRB2017} and treat $g_{c}$ as
an optimization parameter. Using $(\mu, \nu, \gamma)$ as a cyclic permutation of $(x,y,z)$, the exchange
interaction can be written as
\begin{align}
&\sum_{\langle i,j \rangle_\gamma}\sum_{\alpha,\beta = x,y,z}
J_{\alpha\beta}^{\gamma} S_i^{\alpha} S_j^{\beta}\\ \notag
&=\sum_{\langle i,j \rangle_\gamma}\Bigl[ J\,\vec{S}_{i}\cdot\vec{S}_{j} + K S_i^\gamma S_j^\gamma
+ \Gamma\left( S_i^\mu S_j^\nu + S_i^\nu S_j^\mu \right)\\ \notag
&+ \Gamma'\left( S_i^\mu S_j^\gamma + S_i^\nu S_j^\gamma + S_i^\gamma S_j^\mu + S_i^\gamma S_j^\nu \right)\Bigr],
\end{align}
where $J$ is the Heisenberg exchange coupling, $K$ is the Kitaev exchange coupling, $\Gamma$ and
$\Gamma'$ are the off-diagonal exchange couplings. In the following, the parameter set is denoted by
$\bvec{\theta} = (K, \Gamma, \Gamma', J, g_{c})$.

To extract information on low-energy spin degrees of freedom from experimental data, we focus on
the magnetization process of $\alpha$-RuCl$_3$~\cite{Johnson_PRB2015,Kubota_PRB2015}. The magnetic-moment operator is
\begin{align}
\mu_{\eta} = -g_{\eta}\mu_{\rm B}\frac{\vec{h}_{\eta}\cdot\vec{S}}{|\vec{h}_{\eta}|}.
\end{align}
We then obtain the ground state of the Hamiltonian in Eq.~(\ref{eq:ham}), which satisfies the Schr\"odinger
equation,
\begin{align}
\mathcal{H}(\vec{\theta},\vec{h}_{\eta})\Phi_{0}(\vec{\theta},\vec{h}_{\eta})= E_0\Phi_{0}(\vec{\theta},\vec{h}_{\eta}),
\end{align}
to estimate the ground-state magnetization as
\begin{align}
m(\vec{\theta},\vec{h}_{\eta})=\expval{\mu_{\eta}}{\Phi_{0}(\vec{\theta},\vec{h}_{\eta})}.
\label{eq:mag}
\end{align}
where $\Phi_0$ is normalized.

We focus on the magnetization process as the target observable because it directly probes the
spin degrees of freedom and is therefore a natural quantity for constraining the parameters of
spin Hamiltonians. Moreover, the magnetization varies smoothly with the applied field as long as no
first-order phase transition occurs, so that finite-size effects are expected to be small. At sufficiently
low temperatures, the magnetization can be evaluated as a ground-state expectation value via Eq.~(\ref{eq:mag}), which
avoids the computational cost of finite-temperature calculations.

\section{Method}
\label{sec:method}

In the present paper, we perform an unbiased Bayesian analysis of field-direction-dependent magnetization curves,
measured up to about 50~T in Ref.~[\onlinecite{Johnson_PRB2015}]. This analysis complements previous estimates. The
search ranges in Table~\ref{tab:search_range} are the only prior input. Within these ranges, Bayesian optimization automatically
estimates the minimum-cost parameter set. Based on this optimization, we also identify the relevant regions
of the five-dimensional parameter space.

\begin{figure}[tbh]
\begin{center}
\includegraphics[width=0.4\textwidth]{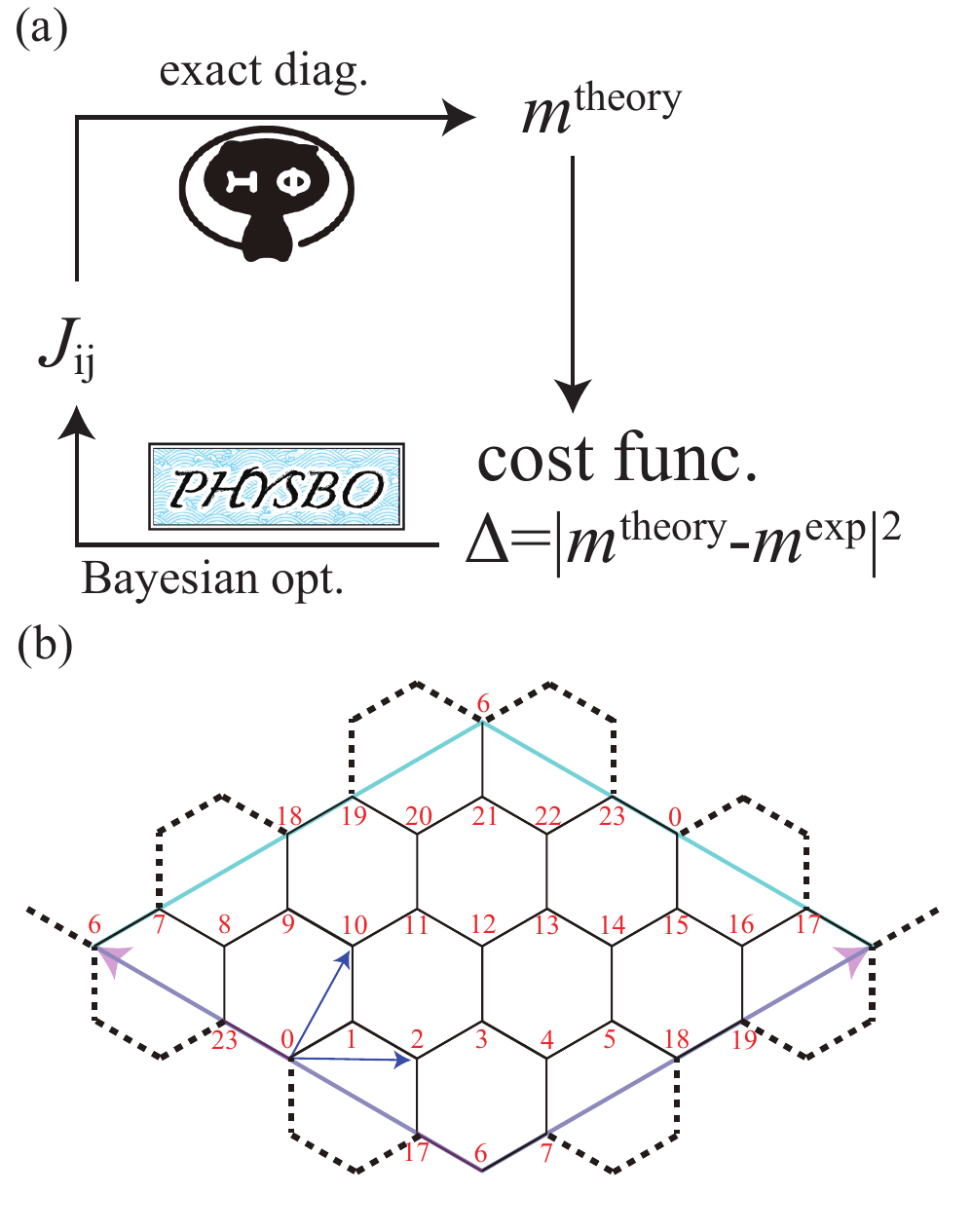}
\end{center}
\caption{(a) Schematic of the Bayesian optimization workflow. For a given parameter set $\bvec{\theta}$, the magnetization
curve is computed by exact diagonalization on a 24-site cluster and compared with experiment via
the cost function $\Delta$. Bayesian optimization then proposes the next candidate, and this cycle is
iterated. (b) 24-site honeycomb cluster with periodic boundary conditions used in this study.
}
\label{fig:schem}
\end{figure}

We employ Bayesian optimization to determine the parameter set $\bvec{\theta}$ that reproduces the experimental magnetization
curves. The workflow is illustrated in Fig.~\ref{fig:schem}. For each candidate parameter set, the magnetization curve
is computed by exact diagonalization and compared with experiment through the cost function $\Delta$ defined
in Eq.~(\ref{eq:delta}). Bayesian optimization then proposes the next candidate, and this cycle is iterated. Each
independent run consists of 500 evaluations (3 random samplings and 497 Bayesian optimizations), and we
perform five independent runs in total.
We restrict the optimization to the nearest-neighbor couplings in Eq.~(\ref{eq:ham}) because the present
five-dimensional optimization is already close to the practical limit of the available computational resources.
In the following subsections, we describe the exact-diagonalization setup,
the cost function, and the Bayesian-optimization procedure.

\subsection{Exact diagonalization and the cost function}
We use the locally optimal block preconditioned conjugate gradient (LOBPCG) method, implemented in $\mathcal{H}\Phi$~\cite{Kawamura_CPC2017,Ido_CPC2024}, to
obtain the ground state of the Hamiltonian in Eq.~(\ref{eq:ham}). All calculations are performed on the
24-site honeycomb cluster shown in Fig.~\ref{fig:schem}(b) with periodic boundary conditions in both directions.
This cluster preserves the rotational symmetry of the honeycomb lattice.

From the experimental data~\cite{Johnson_PRB2015}, measured at 1.4~K, we extract $N_b$ points for the $b$-axis magnetization and $N_c$ points
for the $c$-axis magnetization. In this study, we use $N_b=24$ and $N_c=12$. We digitize the
experimental data using WebPlotDigitizer~\cite{WebPlotDigitizer}.
For each magnetic field, the magnetization is computed by exact diagonalization. The cost function is
defined as
\begin{align}
&\Delta = \frac{1}{N_b + N_c}\Biggl[
\sum_{i=1}^{N_b} \frac{\left( m^{\mathrm{exp}}(\vec{h}_{b}^{\,i}) - m(\vec{\theta},\vec{h}_{b}^{\,i}) \right)^2}{(m_{\mathrm{max}}^b)^2} \notag\\
&+\sum_{i=1}^{N_c} \frac{\left( m^{\mathrm{exp}}(\vec{h}_{c}^{\,i}) - m(\vec{\theta},\vec{h}_{c}^{\,i}) \right)^2}{(m_{\mathrm{max}}^c)^2}
\Biggr],
\label{eq:delta}
\end{align}
where $\vec{h}_{\eta}^{\,i}$ denotes the $i$th magnetic-field vector along the $\eta$-direction, $m^{\mathrm{exp}}(\vec{h}_{\eta}^{\,i})$ is the measured magnetization
at field $\vec{h}_{\eta}^{\,i}$, and $m(\vec{\theta},\vec{h}_{\eta}^{\,i})$ is the theoretical magnetization defined by Eq.~(\ref{eq:mag}).

Here, $m_{\mathrm{max}}^b$ and $m_{\mathrm{max}}^c$ denote the maximum values of the experimental magnetization in each field
direction, used to normalize the squared errors.

The measurement temperature of 1.4~K is far below the dominant exchange scale of the
optimized Hamiltonians obtained below. We also confirmed by the finite-temperature calculations described in Sec.~\ref{sec:chi} that
the magnetization at $\mu_0 H=1$~T for the best parameter set deviates from its low-temperature
limit by less than 0.04\% at 1.4~K for both field directions. These observations
support the ground-state approximation in Eq.~(\ref{eq:mag}).

\subsection{Bayesian optimization}

We perform Bayesian optimization~\cite{Rasmussen_GPML2005} using PHYSBO~\cite{Motoyama_CPC2022}.
PHYSBO constructs a surrogate model of the objective function
by Gaussian process regression and selects the next candidate by maximizing an acquisition function. We
use Thompson sampling~\cite{Russo_FnTML2018} as the acquisition function. In this approach, the Gaussian kernel is approximated
by an $l$-dimensional random Fourier feature vector $\phi(\mathbf{x})$, and the objective function is modeled as
$y = \mathbf{w}^{\top}\phi(\mathbf{x})$, where $\mathbf{w}$ is a weight vector. At each iteration, $\mathbf{w}$ is sampled
from its posterior distribution, and the candidate that maximizes $\mathbf{w}^{\top}\phi(\mathbf{x})$ is selected as the next
evaluation point. Because this procedure requires only $O(l)$ computation per candidate, it scales efficiently to
candidate pools of the present size ($\sim 10^7$). We prepare the candidate pool by combining
parameters within the ranges listed in Table~\ref{tab:search_range}.

\begin{table}[b]
  \centering
  \setlength{\tabcolsep}{9pt}
  \begin{tabular}{cccc}
    \toprule
    Parameter & Range & Step & Number \\
    \midrule
    $K$ (meV) & $[-30,\, 0]$ & $1$ & $31$ \\
    $\Gamma$ (meV) & $[-5,\, 15]$ & $0.5$ & $41$ \\
    $\Gamma'$ (meV) & $[-1,\, 1]$ & $0.1$ & $21$ \\
    $J$ (meV) & $[-10,\, 5]$ & $0.75$ & $21$ \\
    $g_c$ & $[0.5,\, 2.5]$ & $0.1$ & $21$ \\
    \bottomrule
  \end{tabular}
  \caption{Search ranges, step sizes, and number of grid points for the five optimization parameters. The
  total number of candidates is approximately $10^7$.}
  \label{tab:search_range}
\end{table}

Since PHYSBO is designed to maximize an objective function, we use the negative cost function
($-\Delta$ defined by Eq.~(\ref{eq:delta})) as the objective function. Each independent run consists of 3 randomly
selected initial data points, followed by 497 cycles of Bayesian optimization. We perform five independent
runs to derive the spin Hamiltonian.

\begin{table*}[ht]
  \centering
  \setlength{\tabcolsep}{10pt}
  \begin{tabular}{ccccccccc}
    \toprule
    Trial & $K$ & $\Gamma$ & $\Gamma'$ & $J$ & $g_c$ & $\Delta_{\rm opt}$ & $\sqrt{K^2\!+\!\Gamma^2}$ & $T_{\rm peak}$~(K) \\
    \midrule
    No.~1 & $-6.0$  & $10.0$  & $-0.5$ & $-4.75$ & $2.5$ & $0.97\times10^{-3}$ & $11.7$ & $68$\\
    No.~2 & $-6.0$  & $7.5$   & $-0.3$ & $-1.75$ & $2.3$ & $0.91\times10^{-3}$ & $9.6$  & $50$\\
    No.~3 & $-15.0$ & $10.0$  & $-1.0$ & $-6.25$ & $2.5$ & $1.69\times10^{-3}$ & $18.0$ & $89$\\
    No.~4 & $-8.0$  & $11.0$  & $-0.7$ & $-6.25$ & $2.5$ & $1.60\times10^{-3}$ & $13.6$ & $83$\\
    No.~5 & $-14.0$ & $2.0$   & $-0.2$ & $2.0$   & $1.6$ & $2.62\times10^{-3}$ & $14.1$ & $46$\\
    \bottomrule
  \end{tabular}
  \caption{Optimized parameter sets from Bayesian optimization. Energies are given in meV. $\Delta_{\rm opt}$ represents the
  optimized value of the cost function for each trial. $\sqrt{K^2+\Gamma^2}$ is the effective energy scale
  of the Kitaev-$\Gamma$ interaction (in meV), and $T_{\rm peak}$ is the temperature at which the
  specific heat per site $C/\Ns$ exhibits its maximum, obtained from cTPQ calculations on the 24-site
  cluster.}
  \label{tab:results}
\end{table*}

\section{Results}
\label{sec:results}

\begin{figure}[t]
\begin{center}
\includegraphics[width=0.45\textwidth]{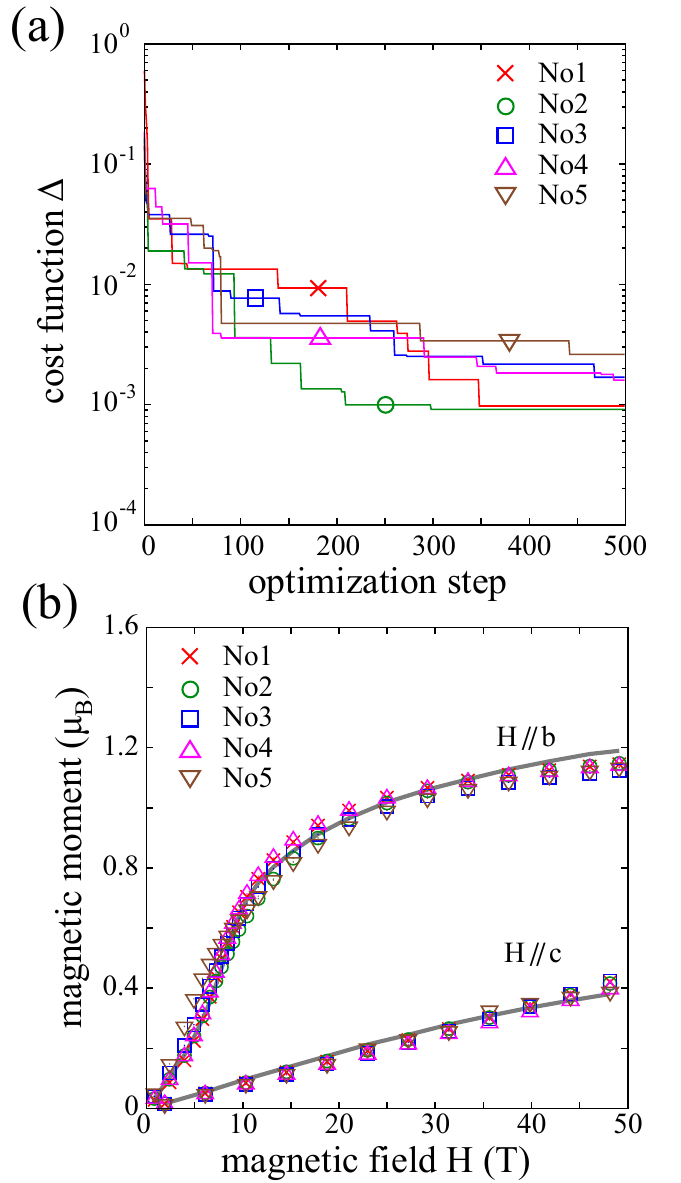}
\end{center}
\caption{(a) Optimization process of the cost function $\Delta$ for five independent runs (No.~1--No.~5). Each run
is independently optimized using 500 evaluations (3 random samplings and 497 Bayesian optimizations). (b) Magnetization
curves along the $b$-axis and the $c$-axis for all five optimized parameter sets (No.~1--No.~5) listed
in Table~\ref{tab:results}, compared with the experimental data (gray curves) from Ref.~[\onlinecite{Johnson_PRB2015}].
All parameter sets reproduce the overall shape of the experimental magnetization process.
}
\label{fig:score}
\end{figure}

\subsection{Bayesian optimization results}
Figure~\ref{fig:score}(a) presents the results of the Bayesian optimization. In all five independent trials, the cost
function $\Delta$ decreases rapidly within the first few hundred steps and subsequently plateaus, confirming convergence
within 500 steps. Since each evaluation requires 36 independent exact-diagonalization calculations on the 24-site cluster
(one per field point along two crystallographic directions), five independent trials of 500 evaluations require
a substantial computational effort. This convergence behavior confirms that five trials adequately sample the cost-function
landscape. Among the five independent runs, No.~2 yields the lowest value of the cost function,
and the corresponding best parameter set is ($K$, $\Gamma$, $\Gamma^{\prime}$, $J$, $g_{c}$) = ($-6.0$, $7.5$,
$-0.3$, $-1.75$, $2.3$). Table~\ref{tab:results} lists the optimal parameter sets from all runs.

Figure~\ref{fig:score}(b) compares the magnetization curves for all five optimized parameter sets with the experimental data.
All parameter sets reproduce the experimental results well in both the $b$- and $c$-axis directions.
In particular, the model correctly captures the rapid increase in magnetization around 5~T along the
$b$-axis and the nearly linear behavior along the $c$-axis. For No.~1--No.~4, the anisotropic field response
is predominantly governed by the large $\Gamma$ term rather than by $g$-factor anisotropy, contrasting with
previous studies~\cite{Yadav_SR2016,Winter_JPCM2017,
Janssen_PRB2017}. No.~5, however, reproduces a comparable magnetization with a small $\Gamma$ and a distinctly
small $g_c = 1.6$, indicating that the magnetization process alone cannot distinguish between these two
mechanisms. As we discuss below, No.~5 does not reproduce other physical quantities well, indicating that
the large-$\Gamma$ scenario is more plausible.

Despite the spread in individual coupling constants, all parameter sets reproduce the overall shape of
the experimental magnetization process, confirming that the Bayesian optimization consistently identifies
physically meaningful regions of the parameter space. Examining Table~\ref{tab:results} in detail,
several common tendencies emerge across the five trials.
The off-diagonal interaction $\Gamma$ is always positive and sizable ($2.0$--$11.0$~meV), while $\Gamma'$ is small and
negative in all cases. The Heisenberg coupling $J$ is negative in four out of five
trials. No.~5 has $J = 2.0$~meV and a distinctly small $g_c = 1.6$, suggesting a
separate local minimum in parameter space. Notably, the two lowest-cost trials (No.~1 and No.~2) share
the same $K = -6.0$~meV but differ in $\Gamma$ and $J$, indicating a degree of
degeneracy in the cost-function landscape.

\begin{figure*}[tb]
\begin{center}
\includegraphics[width=1.0\textwidth]{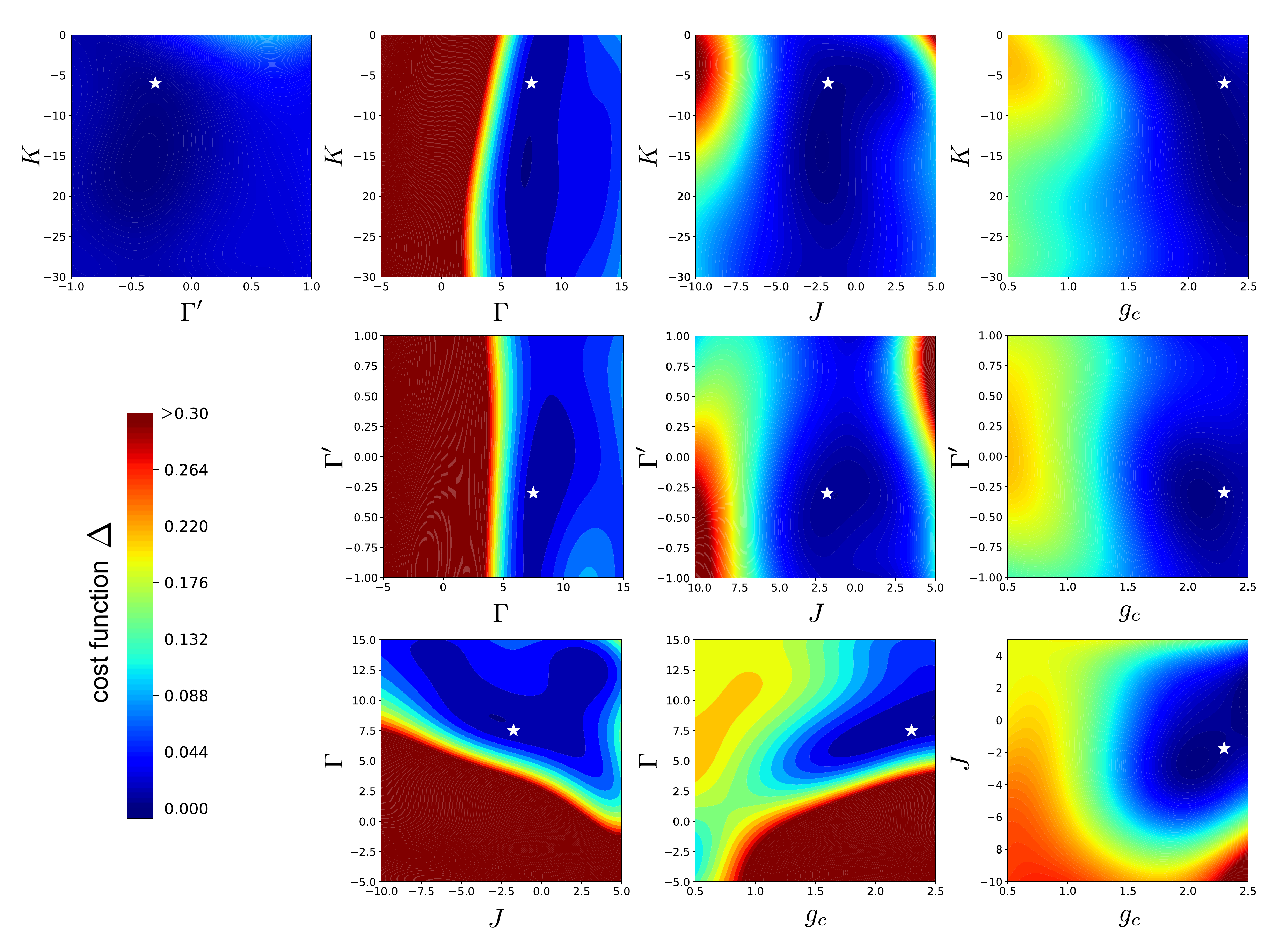}
\end{center}
\caption{Predictive landscapes of the cost function $\Delta$ around the optimal parameters. In each panel, two
parameters are varied while the remaining three are fixed at their optimized values. White stars
indicate the optimized parameters of the lowest-cost run (No.~2).
}
\label{fig:GP_map}
\end{figure*}

\subsection{Cost-function landscape}
\label{sec:landscape}
While several previous studies have proposed parameter sets for the $K$-$J$-$\Gamma$-$\Gamma'$ model of $\alpha$-RuCl$_3$ that
reproduce the magnetization process, few studies have examined the sensitivity of the cost function to
each parameter, i.e., how tightly the magnetization data constrain each coupling constant. This question is
essential, because even if a given parameter set fits the data well, parameters to which
the cost function is insensitive remain essentially undetermined.

To investigate this issue systematically, we construct a surrogate model of the cost function $\Delta$
defined in Eq.~(\ref{eq:delta}) using Gaussian process regression, as implemented in PHYSBO\@. We first train the
Gaussian process on all cost function values accumulated during the Bayesian optimization and then obtain
predictions over the five-dimensional parameter space. Figure~\ref{fig:GP_map} shows the predicted cost function as a function
of two parameters while the remaining three are fixed at their optimized values. Since the
remaining parameters are fixed at the No.~2 values, these cross-sections do not capture local minima
corresponding to other solutions such as No.~5, which lies in a different region of the
five-dimensional parameter space. The Gaussian process is trained on the pooled data from all five
runs and provides a smooth surrogate of the cost function. Hence, the predicted minimum may
deviate slightly from the No.~2 optimal parameters indicated by the white stars. The qualitative structure
of the landscape is nevertheless robust.

\begin{figure*}[t]
\begin{center}
\includegraphics[width=1.0\textwidth]{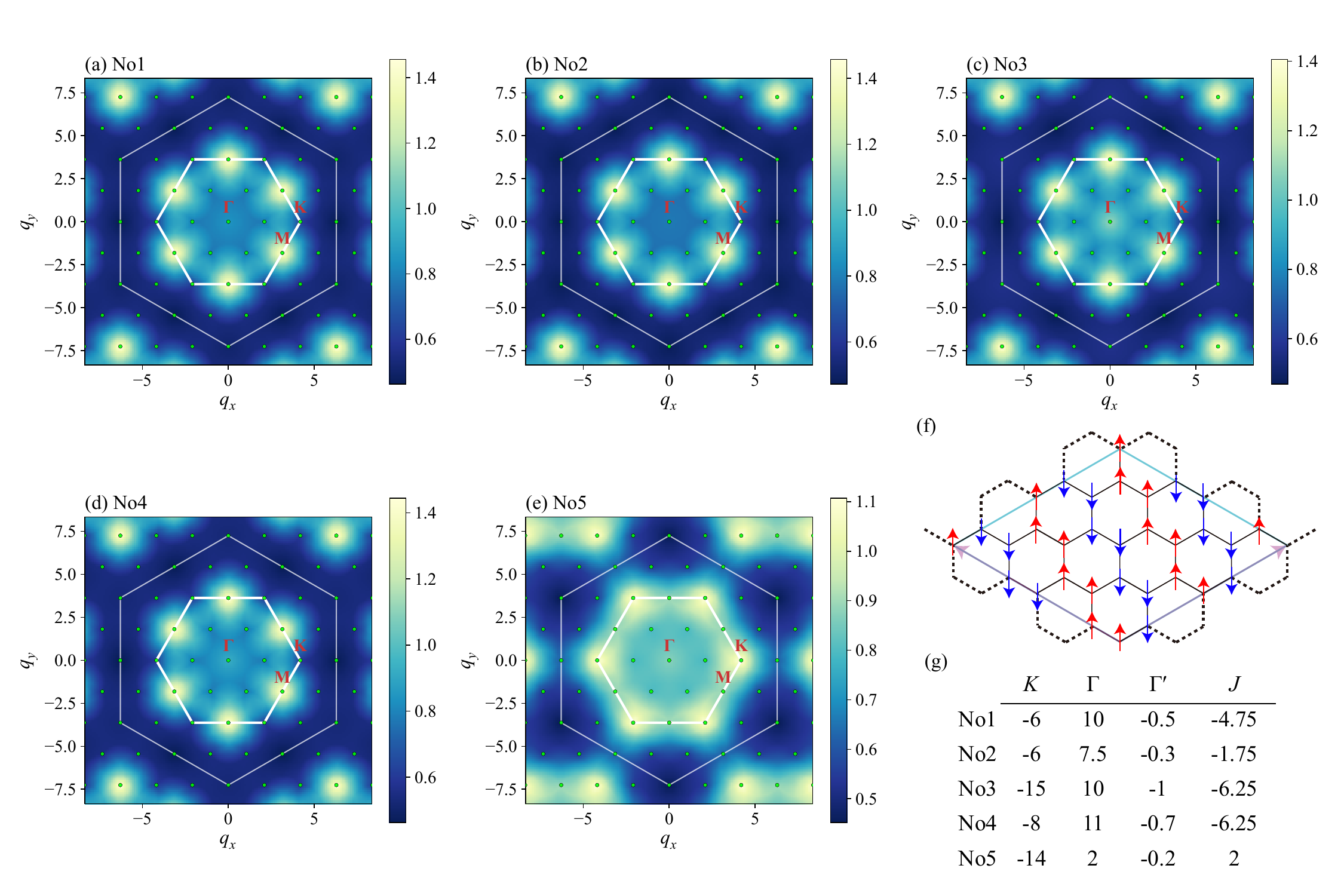}
\end{center}
\caption{Ground-state spin structure factor $S(\mathbf{q})$ in the two-dimensional Brillouin zone for the five optimized parameter
sets (a)--(e) listed in Table~\ref{tab:results}, computed by direct Fourier transform of the real-space spin correlations
at the 12 allowed $\mathbf{q}$ points and continuously interpolated onto a dense $\mathbf{q}$-grid using non-Sibsonian
natural neighbor interpolation~\cite{Belikov_CMMP1997}. The inner (outer) white hexagon denotes the first (extended) Brillouin zone.
Green dots indicate the 12 allowed $\mathbf{q}$ points of the 24-site cluster and their periodic copies.
For No.~1--No.~4, the spectral weight is concentrated at the M points, consistent with zigzag magnetic
order. No.~5 shows a distinct pattern reflecting its different parameter regime.
(f) Schematic of the zigzag magnetic order on the 24-site cluster used in the
exact diagonalization. Red (blue) arrows denote up (down) spins, and the lattice
orientation and the bond labels are the same as in Fig.~\ref{fig:honeycomb}.
The spins are antiparallel only across the $x$ bonds, corresponding to the
$x$ zigzag state discussed in Sec.~\ref{sec:sq}. (g) Exchange couplings
$K$, $\Gamma$, $\Gamma'$, and $J$ (in meV) for No.~1--No.~5.
}
\label{fig:Sq}
\end{figure*}

The maps reveal a clear hierarchy among the five parameters in terms of their influence
on the cost function. The off-diagonal interaction $\Gamma$ exhibits the strongest sensitivity. The cost function
increases sharply when $\Gamma$ is set to negative values, and we observe a well-defined minimum
at $\Gamma > 0$. This indicates that a positive $\Gamma$ is an essential ingredient for
reproducing the experimental magnetization, regardless of the values of the other parameters. The $g$-factor $g_c$
also shows pronounced sensitivity, and the low-cost region is concentrated around $g_c \gtrsim 2.0$. For
$\Gamma'$ and $J$, the cost-function landscape exhibits a moderate sensitivity. While the optimal values tend
to be negative (i.e., $\Gamma' < 0$ and $J < 0$), the cost function remains
relatively flat around the minimum, suggesting that the signs and magnitudes of these parameters are
only weakly constrained by the magnetization process. Indeed, the magnetization process shows weak $\Gamma'$ dependence,
and a small negative $\Gamma'$ is sufficient to reproduce the experimental data. This contrasts with
previous studies that aimed to reproduce high-field ESR spectra, where a large positive $\Gamma'$ was
found to be necessary~\cite{Maksimov_PRR2020,Moller_PRB2025}.

Most notably, the cost function is remarkably insensitive to the magnitude of the Kitaev coupling
$K$. The landscape remains nearly flat over a wide range of $K$ values. This finding
demonstrates that the magnetization process alone cannot determine the absolute value of $K$, which is
a key parameter governing the proximity to the Kitaev spin liquid. This insensitivity is consistent
with the spread of $K$ values observed across the five trials ($K = -6.0$ to
$-15.0$~meV in Table~\ref{tab:results}) and explains why different studies have reported widely varying estimates of $K$
despite achieving comparable fits to the magnetization data.

To identify the weakly constrained direction directly, we examine parameter sets obtained
from No.~2 by changing only $K$. Changing $K$ from $-6$ to $-9$~meV increases the cost function from
$0.91\times10^{-3}$ to $1.52\times10^{-3}$, which remains within the range of the five optimized solutions in
Table~\ref{tab:results}. We present the magnetization and the specific heat for
the systematic $K$ scan in Appendix~\ref{sec:kscan}.

To quantify the uncertainty of the optimized parameters, we estimate a cost-based uncertainty range for each parameter of the
best-fit solution No.~2 by reusing the parameter sets already evaluated during the Bayesian optimization,
without any additional exact-diagonalization calculations. We pool the points evaluated in all five runs,
which include 2449 distinct parameter sets after removing duplicates among the runs. We weight each evaluated point by a Gaussian weight
$w_i=\exp[-(\Delta_i-\Delta_{\rm opt})/(2\sigma^2)]$ and obtain the 95\% range of each parameter from the weighted
quantiles. Here, $\sigma^2$ sets the scale of cost variation regarded as an acceptable fit,
and we adopt $\sigma^2=\Delta_{\rm opt}$, the intrinsic scale of the cost function itself. This choice
is analogous to defining a confidence region from the increase in $\chi^2$ above its
minimum in least-squares fitting. These ranges are not Bayesian credible intervals in the strict sense. The evaluated points are selected by the acquisition function rather than sampled from a posterior distribution. The choice of the weight is also phenomenological. We therefore regard them as approximate ranges of acceptable fits around the optimum. We restrict the analysis to the local basin of No.~2,
defined as the set of evaluated points that lie within a fixed distance of
the optimum in the standardized parameter space and over which the cost increases continuously
from its minimum. Beyond this basin, the cost jumps abruptly as the sampling reaches
the other optimal solutions.

The resulting 95\% cost-based uncertainty ranges are $K \in [-13.0,\, 0.0]$~meV, $\Gamma \in [6.0,\, 11.0]$~meV, $\Gamma' \in
[-0.6,\, 0.1]$~meV, $J \in [-4.0,\, 0.5]$~meV, and $g_c \in [2.1,\, 2.5]$. These ranges show that $\Gamma$, $g_c$,
$J$, and $\Gamma'$ are constrained within finite windows around the best fit, whereas $K$
remains the least constrained parameter. The upper edges of the ranges for $K$ and $g_c$ coincide with
the upper boundaries of the search ranges, and thus these edges are set by the
search ranges rather than by the data. This result quantifies the hierarchy revealed by the landscape
analysis and strengthens our conclusion that the magnetization process alone cannot determine $|K|$.

\subsection{Static spin correlations}
\label{sec:sq}
To examine the validity of our optimized parameter sets, we calculate the ground-state static spin
structure factor in the two-dimensional Brillouin zone. We define the component-resolved structure factor as
\begin{align}
S^{\alpha}(\mathbf{q})=\frac{1}{N}\sum_{i,j} \langle S^{\alpha}_{i} S^{\alpha}_{j} \rangle \, e^{-i\mathbf{q}\cdot(\mathbf{r}_{i}-\mathbf{r}_{j})},
\label{eq:Sq_alpha}
\end{align}
where $\alpha = x, y, z$, $N$ is the total number of sites, and $\mathbf{r}_i$
denotes the real-space coordinate of site $i$ including the sublattice displacement within the unit cell~\cite{Suzuki_PRB2018}.
The total structure factor is given by $S(\mathbf{q}) = \sum_{\alpha} S^{\alpha}(\mathbf{q})$. For the 24-site cluster,
the periodic boundary conditions yield 12 independent $\mathbf{q}$ points in the Brillouin zone. To obtain
a continuous map in momentum space, we evaluate Eq.~(\ref{eq:Sq_alpha}) by direct Fourier transform of the
real-space correlation functions $\langle S^{\alpha}_i S^{\alpha}_j \rangle$ at the 12 allowed $\mathbf{q}$ points and interpolate
the values onto a dense $\mathbf{q}$-grid covering the Brillouin zone
using non-Sibsonian natural neighbor interpolation~\cite{Belikov_CMMP1997}.

Figure~\ref{fig:Sq} shows $S(\mathbf{q})$ computed for all five parameter sets listed in Table~\ref{tab:results}. For No.~1--No.~4, the
spectral weight is concentrated at the M points of the Brillouin zone, indicating zigzag magnetic
order. This is consistent with the experimentally observed zigzag correlations in $\alpha$-RuCl$_3$~\cite{Sears_PRB2015,Banerjee2017}.
No.~5 shows peaks at the K and K$'$ points rather than the M points, inconsistent with the experimental
zigzag order, supporting the large-$\Gamma$ scenario.

Based on the obtained zigzag correlations, we provide an intuitive interpretation of the
insensitivity of the magnetization process to $K$. The
zigzag state consists of ferromagnetic spin chains whose moments alternate from chain to chain,
and only the bonds connecting the up-spin and down-spin chains are antiferromagnetic. Without
loss of generality, we take these antiferromagnetic bonds to be the $x$ bonds [see Fig.~\ref{fig:Sq}(f)]. The
zigzag and polarized states differ only on the $x$ bonds, where the spins change
from antiparallel to parallel, and thus the energy scale of the magnetization process is
governed by the couplings on these bonds. The Kitaev coupling on the $x$ bond
acts only on the $x$ spin component, and its contribution is proportional to $Kn_x^2$,
where $n_x$ is the $x$ component of the ordered-moment direction.

The component-resolved structure factor defined in Eq.~(\ref{eq:Sq_alpha}) shows that $n_x^2$ is small. At the
M point of the zigzag correlations, the spin component along the Kitaev axis of
the antiferromagnetic bonds is strongly suppressed compared with the other two components ($S^{x}(\mathrm{M})=0.26$, while
$S^{y}(\mathrm{M})=S^{z}(\mathrm{M})=0.62$ for No.~2). This smallness of $n_x^2$ provides an intuitive explanation of why the magnetization
process is almost insensitive to $K$. In contrast, the Heisenberg coupling acts on all
spin components and contributes with its full weight.
We also give a complementary mean-field argument for the classical spin model in Appendix~\ref{sec:mf_slope}.

\subsection{Magnetic susceptibility}
\label{sec:chi}
\begin{figure}[tb]
\begin{center}
\includegraphics[width=0.45\textwidth]{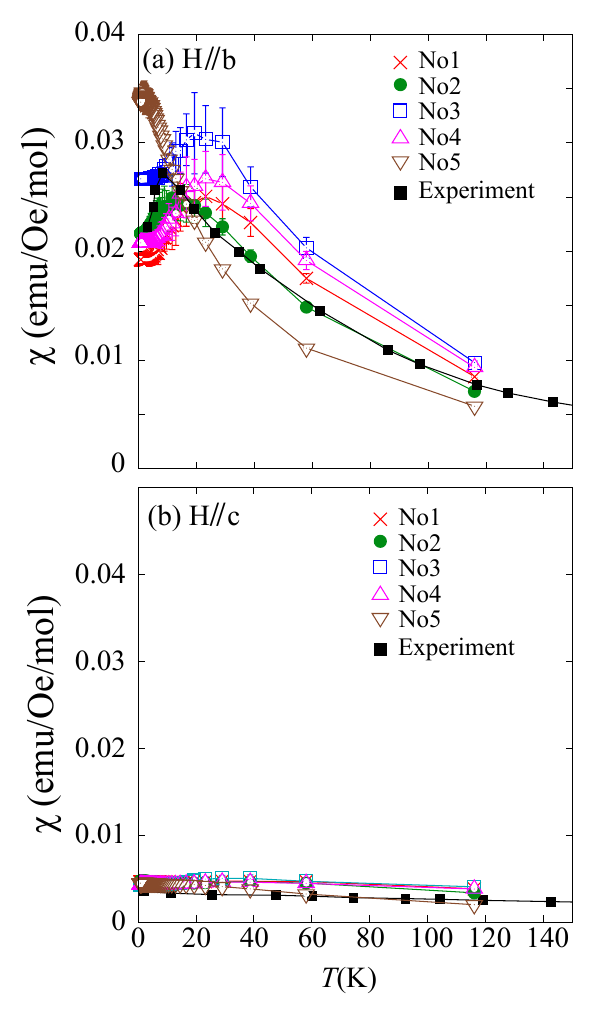}
\end{center}
\caption{Temperature dependence of the magnetic susceptibility $\chi = M/H$ at $H=1$~T calculated by the cTPQ
method for (a)~$H\perp c$ ($b$-axis) and (b)~$H\parallel c$ ($c$-axis). We show experimental data from Ref.~[\onlinecite{Weber2016}]
for comparison. All parameter sets reproduce the strong anisotropy $\chi_b \gg \chi_c$.
}
\label{fig:chi}
\end{figure}

We evaluate finite-temperature properties of our optimized models using the canonical thermal pure quantum (cTPQ)
state method~\cite{Sugiura_PRL2012,Sugiura_PRL2013} on the same 24-site cluster. The unit conversion between computed and experimental quantities
is summarized in Appendix~\ref{sec:unit_conversion}. We compute the magnetic susceptibility $\chi(T) = M/H$ at $H=1$~T using
the cTPQ method for all parameter sets. Figure~\ref{fig:chi} shows $\chi(T)$ along the $b$-axis ($H\perp c$)
and $c$-axis ($H\parallel c$), together with experimental data from Ref.~[\onlinecite{Weber2016}] digitized using WebPlotDigitizer~\cite{WebPlotDigitizer}. The strong
susceptibility anisotropy has also been discussed experimentally in Ref.~\cite{LampenKelley2018}. All parameter sets reproduce the strong
anisotropy $\chi_b \gg \chi_c$ observed experimentally.

Along the $b$-axis [Fig.~\ref{fig:chi}(a)], No.~2 reproduces the experimental susceptibility remarkably well,
indicating that this parameter
set captures not only the ground-state magnetization but also finite-temperature magnetic excitations. No.~1, No.~3, and
No.~4 also show reasonable agreement, although the peak positions and magnitudes deviate slightly from the
experimental data. No.~5, however, fails to reproduce the $b$-axis susceptibility even qualitatively, further supporting the
large-$\Gamma$ scenario (No.~1--No.~4).

Along the $c$-axis [Fig.~\ref{fig:chi}(b)], the susceptibility is much smaller and nearly temperature-independent above $\sim 20$~K,
consistent with the experimental observations. The differences among parameter sets are less pronounced along the
$c$-axis, indicating that the $c$-axis susceptibility is less sensitive to the exchange parameters than the
$b$-axis susceptibility.

\subsection{Specific heat}
\label{sec:specific_heat}
\begin{figure}[tb]
\begin{center}
\includegraphics[width=0.45\textwidth]{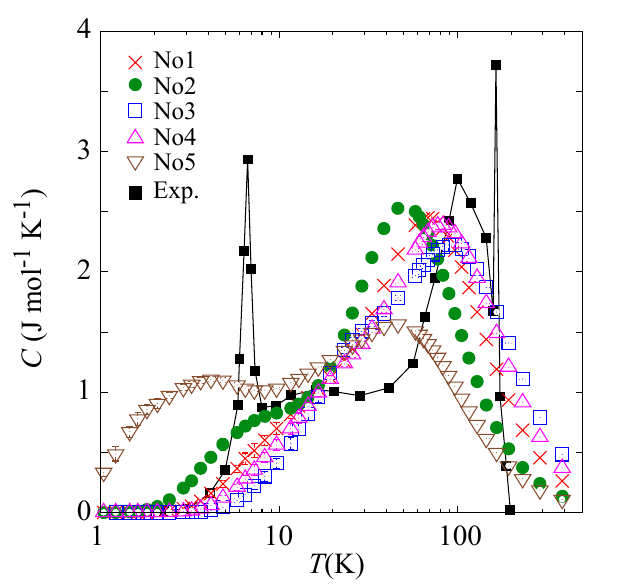}
\end{center}
\caption{Temperature dependence of the specific heat per site, $C/\Ns$, calculated by the cTPQ method on
the 24-site cluster for the parameter sets listed in Table~\ref{tab:results}, together with experimental data from
Ref.~[\onlinecite{Do_NPhys2017}]. For No.~1--No.~4, the peak temperature $T_{\rm peak}$ increases with the effective energy scale $\sqrt{K^2+\Gamma^2}$,
while No.~5 deviates from this trend.
}
\label{fig:specific_heat}
\end{figure}

We also compute the temperature dependence of the specific heat per site $C/\Ns$, as shown
in Fig.~\ref{fig:specific_heat}, together with experimental data from Ref.~[\onlinecite{Do_NPhys2017}] digitized using WebPlotDigitizer~\cite{WebPlotDigitizer}. As listed in Table~\ref{tab:results},
the peak temperature $T_{\rm peak}$ of the specific heat varies substantially across the parameter sets,
ranging from $\sim 46$~K (No.~5) to $\sim 89$~K (No.~3). In trials No.~1--No.~4, $T_{\rm peak}$ tends
to increase with the effective energy scale $\sqrt{K^2+\Gamma^2}$, whereas No.~5 deviates from this trend. This
suggests that the Kitaev--$\Gamma$ energy scale is an important, but not exclusive, factor governing the
position of the specific-heat maximum.

The best parameter set (No.~2), which yields the lowest cost function for the magnetization, gives
$T_{\rm peak}\approx 50$~K. However, the position of the high-temperature peak deviates significantly. Reproducing the high-temperature
peak requires a substantially larger interaction energy scale~\cite{Suzuki_PRB2018,Laurell_npjQM2020,
Li_NC2021}. A previous study has subtracted the phonon
contribution from the experimental specific heat of $\alpha$-RuCl$_3$ using a nonmagnetic analog RhCl$_3$ and $ab$
$initio$ phonon calculations. As a result, it is found that a broad excess peak persists
around $\sim 70$~K even after the subtraction~\cite{Widmann_PRB2019}. This suggests that the high-temperature peak has a
magnetic origin. As shown in Sec.~\ref{sec:landscape}, the magnetization process is insensitive to $|K|$, so that
larger values of $|K|$ are not excluded by the magnetization fitting.

To demonstrate this point directly, we perform a $K$-only scan in which $K$ is
changed from $-6$ down to $-15$~meV with the other No.~2 parameters fixed. The
magnetization and specific-heat curves along this scan are presented in Appendix~\ref{sec:kscan}. The
specific-heat maximum shifts systematically from $50.4$ to $72.5$~K and approaches the broad
experimental feature around $\sim70$~K, while the magnetization fit degrades only gradually. For the
representative case $K=-9$~meV, which increases $\sqrt{K^2+\Gamma^2}$ from $9.6$ to $11.7$~meV, the cost function
($1.52\times10^{-3}$) remains within the range of the five optimized solutions in Table~\ref{tab:results}. The static
spin structure factor of this set retains a clear maximum at the M point
($S(\mathrm{M})=1.47$, compared with $1.50$ for No.~2). Increasing $|K|$ therefore improves the position of the
high-temperature specific-heat peak without substantially degrading the magnetization fit or changing the dominant
zigzag correlation.

Experimentally, the specific heat of $\alpha$-RuCl$_3$ exhibits a sharp peak at the N\'{e}el temperature $T_{\rm
N}\approx 7$~K~\cite{Sears_PRB2015}, which our 24-site two-dimensional cluster calculation cannot reproduce due to the finite-size limitation.
Nevertheless, No.~2 shows a shoulder structure around $\sim 10$~K, which we attribute to the entropy
release associated with the development of magnetic order. This feature is qualitatively consistent with the
experimental low-temperature peak. As in the case of the magnetic susceptibility, No.~5 shows qualitatively different
behavior from No.~1--No.~4, further supporting the large-$\Gamma$ scenario.

This observation has an important implication for the inverse problem of determining spin Hamiltonian parameters
from experimental data. While the magnetization process strongly constrains the ratios among exchange couplings and
the $g$-factor, it is relatively insensitive to the overall energy scale $\sqrt{K^2+\Gamma^2}$. The cost functions
for No.~1 ($\Delta_{\rm opt}=0.97\times10^{-3}$, $\sqrt{K^2+\Gamma^2}=11.7$~meV) and No.~2 ($\Delta_{\rm opt}=0.91\times10^{-3}$, $\sqrt{K^2+\Gamma^2}=9.6$~meV) are nearly identical despite differing
energy scales, whereas their specific heat peaks differ by $\sim 20$~K. This demonstrates that magnetization
data alone cannot uniquely determine the absolute energy scale of exchange interactions. Temperature-dependent quantities such
as the specific heat can in principle lift this degeneracy, but it requires careful treatment
of non-spin contributions such as phonons and spin-phonon coupling at high temperatures.

\section{Summary}
\label{sec:summary}

In this paper, we have presented a Bayesian optimization approach to determine the spin Hamiltonian
parameters of $\alpha$-RuCl$_3$ from experimental magnetization process data. By optimizing five parameters (the Kitaev interaction
$K$, the off-diagonal interactions $\Gamma$ and $\Gamma'$, the Heisenberg interaction $J$, and the $g$-factor $g_c$),
we obtained the best parameter set $(K, \Gamma, \Gamma', J, g_c) = (-6.0, 7.5, -0.3,
-1.75, 2.3)$ (exchange couplings in meV), which quantitatively reproduces the experimental magnetization curves along both the $b$-
and $c$-axis directions. The magnetization process can be reproduced either by a large $\Gamma$ with
$g_c \gtrsim 2.0$ (No.~1--No.~4) or by a small $\Gamma$ with a small $g_c$ (No.~5), so
that the magnetization data alone cannot distinguish between these two scenarios.

By analyzing the cost-function landscape using Gaussian process regression, we found that the cost function
is most sensitive to $\Gamma$ and $g_c$, moderately sensitive to $J$ and $\Gamma'$, and least
sensitive to $K$. In particular, the insensitivity to $K$ implies that the magnetization process alone
cannot determine the absolute energy scale of exchange interactions.

We examined additional physical quantities to discriminate between the two scenarios. The ground-state spin structure
factor $S(\mathbf{q})$ shows clear peaks at the M points for No.~1--No.~4, indicating zigzag magnetic order
consistent with experiment, while No.~5 exhibits peaks at the K and K$'$ points, inconsistent with
the observed zigzag order. The magnetic susceptibility $\chi(T)$ computed by the cTPQ method further supports
this conclusion. No.~2 reproduces the experimental $b$-axis susceptibility remarkably well, whereas No.~5 fails qualitatively. The
specific heat also shows qualitatively different behavior for No.~5. These results consistently favor the large-$\Gamma$
scenario~\cite{Kim_PRB2016,Winter_PRB2016,
Ran_PRL2017,LampenKelley2018}. Reproducing the experimental high-temperature peak in the specific heat requires a larger interaction energy scale, and
temperature-dependent quantities can in principle constrain this energy scale, but it requires careful treatment of
non-spin contributions such as phonons.

Our results indicate that $\Gamma$ is sizable and cannot be neglected relative to $|K|$
in $\alpha$-RuCl$_3$, and thus the material is not in close proximity to the pure
Kitaev limit with $\Gamma \simeq 0$. Since a
large positive $\Gamma$ competes with and can destabilize the Kitaev spin liquid, the realization
of the pristine Kitaev spin liquid in $\alpha$-RuCl$_3$ is not straightforward. We note that
the magnetization process leaves $|K|$ essentially undetermined, so that a sizable Kitaev coupling
is not excluded by our analysis. The $\Gamma$-dominant regime is also strongly frustrated, and
its possible quantum-disordered states have
been actively discussed~\cite{Rousochatzakis_PRL2017,Catuneanu_npjQM2018,PhysRevB.97.075126,Luo_npjQM2021,Rousochatzakis_RPP2024}. The fate of the field-induced disordered regime of $\alpha$-RuCl$_3$ may
need to be reconsidered from the viewpoint of $\Gamma$-dominated physics.

The combination of Bayesian optimization with accurate low-energy solvers such as exact diagonalization provides a
systematic and unbiased framework for determining spin Hamiltonian parameters from experimental data. Since the method
requires only the evaluation of a cost function for each candidate parameter set, it is
readily applicable to other Kitaev materials~\cite{Winter_JPCM2017} and quantum magnets where the determination of effective spin
models remains challenging.
For $\alpha$-RuCl$_3$ itself, a promising next step is to fix $g_c$ at the value
determined here and to extend the search space to further-neighbor couplings such as $J_3$.
The resolving power of the method depends on the information content of the target
data. For a material whose magnetization curves are comparatively featureless, such as Na$_2$IrO$_3$~\cite{PhysRevB.82.064412}, the
cost-function landscape becomes flatter and the constraints on the exchange parameters weaken. In
such cases, we can perform a similar analysis by targeting other observables with
characteristic features, such as the anisotropic susceptibility or the specific heat. This extension
is an interesting direction for future studies.

\acknowledgments
This research was supported by JSPS KAKENHI (Grant Nos. JP20H01850, JP23H03818, and JP26K00652). TM was supported
by JST FOREST JPMJFR236N. YY was supported by JSPS KAKENHI (Grant Nos. 22H01183, 23H04524, and
25K07233). This research was also partly supported by MEXT as ``Basic Science for Emergence and
Functionality in Quantum Matter -- Innovative Strongly-Correlated Electron Science by Integration of Fugaku and Frontier
Experiments'' (JPMXP1020200104) as a program for promoting research on the supercomputer Fugaku, supported by RIKEN-Center
for Computational Science (R-CCS) through HPCI System Research Project (Project ID: hp200132, hp210163, and hp220166).
Numerical calculations were performed using the facilities of the Supercomputer Center, The Institute for Solid
State Physics, The University of Tokyo.
The numerical data, input files, and analysis scripts used in this study are available in the ISSP Data
Repository~\cite{rucl3_bo_mag_data}.

\appendix

\section{Unit conversion for magnetic susceptibility and specific heat}
\label{sec:unit_conversion}
In this Appendix, we summarize how to convert the computed physical quantities into the units
used in experimental measurements. Table~\ref{tab:constants} lists the physical constants used in the conversion. We denote
by $n$ the number of magnetic ions per formula unit ($n=1$ for $\alpha$-RuCl$_3$).

\begin{table}[h]
  \centering
  \setlength{\tabcolsep}{5pt}
  \begin{tabular}{lll}
    \toprule
    Name & Symbol & Value \\
    \midrule
    Avogadro number & $N_{\rm A}$ & $6.02214 \times 10^{23}$~mol$^{-1}$ \\
    Bohr magneton & $\mu_{\rm B}$ & $9.27401 \times 10^{-24}$~J/T \\
    Boltzmann constant & $k_{\rm B}$ & $1.38065 \times 10^{-23}$~J/K \\
    Gas constant & $R = N_{\rm A} k_{\rm B}$ & $8.31446$~J\,mol$^{-1}$\,K$^{-1}$ \\
    ---        & $1$~meV & $1.60218 \times 10^{-22}$~J \\
    \bottomrule
  \end{tabular}
  \caption{Physical constants used in the unit conversion.}
  \label{tab:constants}
\end{table}

\textbf{Temperature axis}: When exchange couplings are given in meV, the temperature axis is expressed
as $k_{\rm B}T$. The conversion to kelvin is
\begin{align}
T~[\mathrm{K}] = \frac{1}{k_{\rm B}} \times k_{\rm B}T = 11.6045 \times k_{\rm B}T,
\end{align}
where $k_{\rm B}T$ is measured in meV.

\textbf{Magnetic susceptibility}: The computed susceptibility per spin, $\chi$ [$\mu_{\rm B}$/T], is converted to the molar
susceptibility $\chi_M$ [emu\,mol$^{-1}$\,Oe$^{-1}$] by
\begin{align}
\chi_M = n \times N_{\rm A} \mu_{\rm B} \times 10^{-1} \times \chi = 0.5585\, n \times \chi,
\end{align}
where $N_{\rm A} \mu_{\rm B} = 5.5849$~J\,T$^{-1}$\,mol$^{-1}$, and the factor $10^{-1}$ arises from $1~\mathrm{emu} =
10^{-3}~\mathrm{J/T}$ and $1~\mathrm{T} = 10^{4}~\mathrm{Oe}$.

\textbf{Specific heat}: The computed specific heat per spin, $c$ [$k_{\rm B}$], is converted to the
molar specific heat $C_{\mathrm{mol}}$ [J\,mol$^{-1}$\,K$^{-1}$] by
\begin{align}
C_{\mathrm{mol}} = nR \times c.
\end{align}

\section{$K$ dependence of the magnetization and specific heat}
\label{sec:kscan}
\begin{figure}[tb]
\begin{center}
\includegraphics[width=0.42\textwidth]{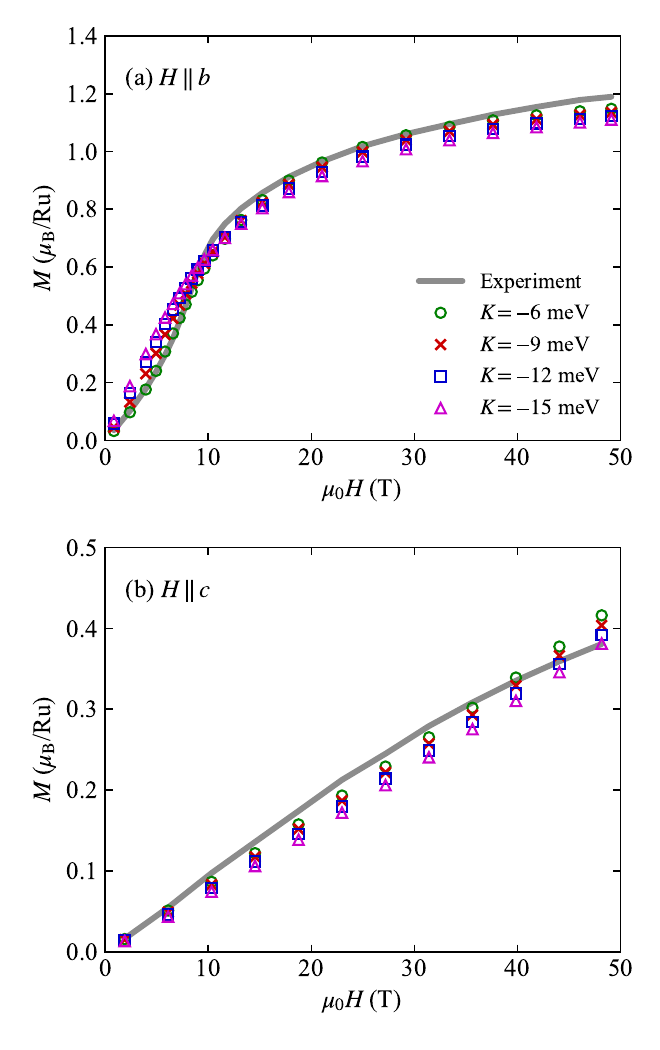}
\end{center}
\caption{Magnetization curves for (a) $H\parallel b$ and (b) $H\parallel c$ obtained by changing only
$K$ in the No.~2 parameter set ($\Gamma=7.5$~meV, $\Gamma'=-0.3$~meV, $J=-1.75$~meV, $g_c=2.3$). The gray curves denote
the experimental data from Ref.~[\onlinecite{Johnson_PRB2015}].}
\label{fig:mag_kscan}
\end{figure}

\begin{figure}[tb]
\begin{center}
\includegraphics[width=0.45\textwidth]{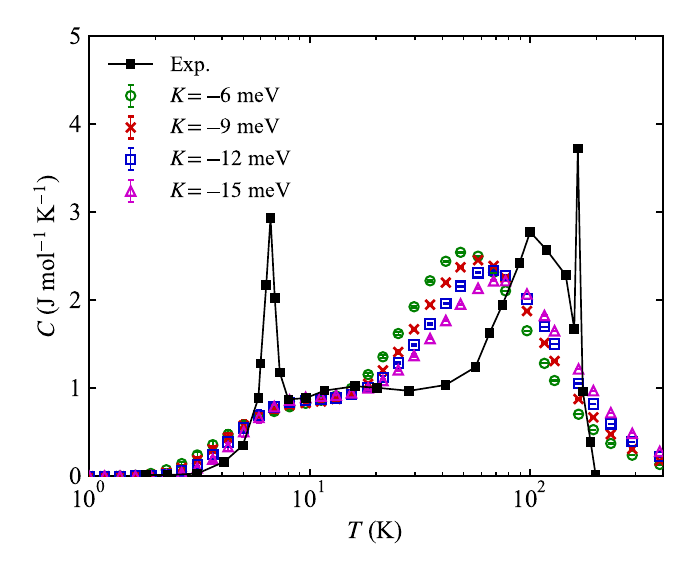}
\end{center}
\caption{Temperature dependence of the specific heat for the same $K$-only scan as in
Fig.~\ref{fig:mag_kscan}, calculated by the cTPQ method on the 24-site cluster with 72 random samples,
together with the experimental data from Ref.~[\onlinecite{Do_NPhys2017}].}
\label{fig:C_kscan}
\end{figure}

In this Appendix, we present the magnetization and the specific heat for the parameter
sets obtained from No.~2 by changing only $K$. Figure~\ref{fig:mag_kscan} compares the calculated magnetization curves
for $K=-6$, $-9$, $-12$, and $-15$~meV with the experimental data. The cost function increases monotonically
with $|K|$ ($10^3\times\Delta=0.91$, $1.52$, $2.81$, and $4.41$, respectively), while the calculated curves retain the overall
agreement with experiment. Figure~\ref{fig:C_kscan} shows the corresponding cTPQ specific heat. The high-temperature maximum shifts
systematically from $50.4$ to $72.5$~K with increasing $|K|$ and approaches the broad experimental feature
around $\sim70$~K. These results demonstrate that increasing $|K|$ improves the position of the high-temperature specific-heat
peak while the cost function gradually becomes worse. We therefore
adopt $K=-9$~meV as the representative set discussed in the main text.

\section{Mean-field analysis of the initial slope of the magnetization}
\label{sec:mf_slope}
In this Appendix, we give a mean-field argument for the classical spin model.
This argument offers a simple explanation of why the magnetization process is insensitive to $K$,
which supplements the discussion in Sec.~\ref{sec:sq}.
For simplicity, we assume that the ground state is the zigzag order.
In this analysis, we ignore $\Gamma'$ because its amplitude is small.
In the zigzag order, there are two types of chains. To label the zigzag order on the present honeycomb lattice,
we use the bond label to specify one of the three possible zigzag patterns.
For example, in the $x$ zigzag state, spins are ferromagnetically aligned along chains consisting of
alternating $y$ and $z$ bonds, while spins on neighboring chains are antiferromagnetically aligned across the $x$ bonds [see Fig.~\ref{fig:Sq}(f)].
Here, we assume the ordered moment $+\vec{m}$ on one type of chain and $-\vec{m}$ on the other.
When external magnetic fields $\vec{h}$ are applied, the induced uniform moment $\vec{\delta}$ is assumed to modify the magnetic moments
on these chains as $\pm\vec{m}+\vec{\delta}$.

When we assume that $\left|\pm\vec{m}+\vec{\delta}\right|^2 = s^2$, where $s$ is the saturated moment per site,
the mean-field energy per $x$ bond, i.e., per two sites, $\epsilon$, is given as
\begin{eqnarray}
\epsilon
&=&
K|\vec{m}|^2 - 2 K m_x^2 + K|\vec{\delta}|^2 + J |\vec{m}|^2 + 3J |\vec{\delta}|^2
\nonumber\\
&&+
2\Gamma \left( -m_y m_z + m_z m_x + m_x m_y  \right)
\nonumber\\
&&+
2 \Gamma \left( \delta_y \delta_z + \delta_z \delta_x + \delta_x \delta_y \right)
- 2\vec{h}\cdot \vec{\delta}.
\label{eq:mf_ene0}
\end{eqnarray}
The condition $\left|\pm\vec{m}+\vec{\delta}\right|^2 = s^2$ requires $\vec{m}\cdot\vec{\delta}=0$.
The terms depending on $\vec{m}$ can be written as $(J+K)|\vec{m}|^2 + \vec{m}^{T} A \vec{m}/2$ with
\begin{eqnarray}
A=
\left(
\begin{array}{ccc}
-4K & +2\Gamma & +2\Gamma \\
+2\Gamma & 0 & -2\Gamma \\
+2\Gamma & -2\Gamma &  0 \\
\end{array}
\right).
\label{eq:mf_A}
\end{eqnarray}
When we set $\vec{\delta}=\vec{0}$ and $\vec{h}=\vec{0}$, the minimization of $\epsilon$ under the constraint
$|\vec{m}|^2=s^2$ shows that $\vec{m}$ is parallel to the eigenvector $\bvec{n}_1$ of $A$ with the lowest eigenvalue $\lambda_1$.
For finite $\vec{\delta}$, we keep the direction of $\vec{m}$ and set
$\vec{m} = \sqrt{s^2-|\vec{\delta}|^2}\bvec{n}_1$. We then
obtain the energy per $x$ bond as
\begin{eqnarray}
\epsilon
&=&
(J+K+\lambda_1 / 2)s^2
+(2J-\lambda_1 /2 - \Gamma)|\vec{\delta}|^2
\nonumber\\
&&+ \Gamma (\delta_x + \delta_y + \delta_z)^2
- 2\vec{h}\cdot \vec{\delta}.\label{eq:mf_ene}
\end{eqnarray}

Then, the energy per $x$ bond in Eq.~(\ref{eq:mf_ene}) provides the mean-field description
of the uniform magnetization $\vec{\delta}$ under the applied magnetic field $\vec{h}$.
While the formula Eq.~(\ref{eq:mf_ene}) describes easy-plane anisotropy for $\Gamma > 0$,
the coefficient of the term proportional to $|\vec{\delta}|^2$
determines the initial slope of the magnetization process.
The easy-plane anisotropy mainly comes from the term $\Gamma(\delta_x+\delta_y+\delta_z)^2$.
For the field along the $b$-axis, this term vanishes because $\delta_x+\delta_y+\delta_z=0$ is satisfied.
On the other hand, it becomes $3\Gamma|\vec{\delta}|^2$ for the field along the $c$-axis.
Thus, for $\Gamma>0$, this term gives the smaller initial slope along the $c$-axis than along the $b$-axis.

Here, we examine how
the coefficient $(2J-\lambda_1/2 -\Gamma)$
depends on the exchange couplings.
The Heisenberg coupling $J$ enters the coefficient directly as $2J$.
In contrast, the Kitaev coupling $K$ affects the coefficient only through $\lambda_1$.
For $\Gamma > 0$, the $K$ dependence of the lowest eigenvalue of $A$, $\lambda_1$, is given by
\begin{eqnarray}
\lambda_1 =
-(\Gamma + 2K)-\sqrt{4K^2-4K\Gamma+9\Gamma^2},
\label{eq:mf_lambda1}
\end{eqnarray}
which, for $K<0$,
satisfies
\begin{eqnarray}
2J < 2J
-\lambda_1 /2 - \Gamma
< 2J + \Gamma.
\label{eq:mf_bound}
\end{eqnarray}
Equation~(\ref{eq:mf_bound}) shows that the $K$-dependent part of the coefficient varies only
within a window of width $\Gamma$, and the coefficient is mainly governed by $J$ and $\Gamma$.
Therefore, the initial slope of the magnetization is insensitive to $K$.
This suggests why the magnetization process does not constrain $K$.

\end{document}